\documentclass[prb,twocolumn]{revtex4}
\usepackage{epsfig}
\begin{document}

\title{Schr\"odinger equation with a spatially and temporally random
potential: Effects of cross-phase modulation in optical
communication.}
\author{A. G. Green$^1$, P. B. Littlewood$^2$, P. P. Mitra$^3$ and L. G. L.
Wegener$^2$}
\affiliation{
$^1$Theoretical Physics, Oxford University, U.K.\\
$^2$T.C.M., Cavendish Laboratory, Cambridge University, UK.\\
$^3$Lucent Technologies, Bell Laboratories Innovations, Murray Hill, NJ.}
\date{\today}

\begin{abstract}
We model the effects of cross-phase modulation in frequency (or wavelength)
division multiplexed optical communications systems, using a
Schr\"odinger equation with a spatially and temporally random
potential. Green's functions for the propagation of light in this
system are calculated using Feynman path-integral and diagrammatic
techniques. This propagation leads to a non-Gaussian joint
distribution of the input and output optical fields. We use these
results to determine the amplitude and timing jitter of a signal pulse
and to estimate the system capacity in analog communication.
\end{abstract}
\maketitle

\section{Introduction}

\subsection{General}

The ability to transmit information is ultimately limited by signal
distortion. Information theory\cite{Shannon48,Cover91} quantifies the
extent to which such distortions inhibit communication; it was
originally developed for the study of radio communication or
electrical communication along copper wires. In these
cases the signal propagation is linear- the received signal is a linear
function of the transmitted signal. Distortion arises due to the
addition of extraneous signal fluctuations, which propagate
linearly alongside the original signal. These fluctuations may
come from noisy amplifiers or other circuit elements, or from
cross-talk with other messages. Such linear systems with additive
noise are very well characterized from an information theoretical
perspective.

In modern communication systems, the situation is rather more
complicated. The very high transmission rates- particularly in
optical communication- require that the transmission medium be
operated in regimes where the signal propagation is substantially
non-linear\cite{Agrawal95}. These nonlinearities can lead to a variety of new
mechanisms of signal
distortion\cite{Gordon86,Gordon90,Mollenauer91,Marcuse90,Humblet91,Hui97}
and are   
often very difficult to characterize analytically.  This has limited
the understanding of non-linear channels, particularly from an
information theoretical perspective.

Faced with these difficulties, there are two ways in which
progress may be made. One approach is to perform detailed
numerical simulations of the underlying partial differential
equation. This has been the approach of much of the work in the
literature. However, it is very difficult to use the results of
such analyses to understand the information theoretical aspects of
communication in non-linear media. An alternative approach is to
harness physical understanding of the signal propagation to
motivate simple phenomenological models that capture some aspects
of the non-linear propagation, but which are nevertheless
analytically tractable. This approach has been used to understand
a number of aspects of non-linear propagation in optical fibers
\cite{{Gordon86,Gordon90,Mollenauer91,Marcuse90,Humblet91,Hui97}}. Only
recently have such 
approaches been applied to study the information theoretical
aspects of the problem\cite{Mitra01,Narimanov02,Mitra02}.

When one is interested in optimizing the design of a
communications system, it is often useful to characterize the
various mechanisms of signal distortion according to their physical
origin in the system. This is not always the best way of classifying
signal distortion from an information theoretical point of view. A
more convenient classification is to divide signal distortions
into additive and multiplicative noise. The signal
distortions in optical fibers may be classified in this
manner\cite{Mitra02}. It is the multiplicative noise that presents
the main difficulty in understanding non-linear channels. It lies
behind some of the classic unsolved problems in information
theory\cite{Cover91}. It is to this aspect of the problem that
path-integral and diagrammatic techniques are particularly suited.
In the system studied in this paper, multiplicative noise appears as a
random potential in a Schr\"odinger equation.

The tension between additive and multiplicative noise is similar to
the tension between disorder and interactions in many-body physics. In
the latter case, the consequence of this tension is that a full
analytical characterization is not possible. Instead, one must resort
to approximation schemes that are appropriate under different regimes
of system parameters. It is likely that for the same reasons one will
be forced to use similar approximate schemes to study the propagation
of signals in non-linear media. Physicists have worked hard to develop
such schemes and some of 
them are directly applicable to the study of non-linear signal
propagation; for example, Falkovich et al\cite{Falkovich01} have used
the theory of optimal fluctuation to understand the fluctuations
in soliton amplitude and timing that occur in optical
communication. In this paper, we use Feynman path-integral\cite{Feynman} and
diagrammatic techniques\cite{Efetov} in order to understand some of the
effects of non-linearity upon optical communication. Most of this paper will
concentrate upon characterizing the propagation of light in the
fiber. Towards the end of the paper, we will harness this
characterization to understand some information theoretical
aspects of the problem.

\subsection{The wavelength division multiplexed (WDM) optical fiber.}

In this paper, we will consider some of the effects of
non-linearity upon communication using frequency division
multiplexed optical fibers\cite{Agrawal95}. Optical fibers have an enormous
transparent bandwidth. It is far beyond current electronics to
modulate at these frequencies. In order to utilize as much of the
bandwidth as possible, it is divided into a number of
non-overlapping frequency bands, or sub-bands, each of which is
modulated independently. This is called frequency division
multiplexing or alternatively wavelength division
multiplexing(WDM).

Light propagating in a WDM optical fiber experiences several
sources of signal distortion and non-linearity. The first of these
is due to dispersion; the different frequency components of a
signal travel at different velocities along the fiber, leading to a
spreading of a signal pulse. This spreading is a linear operation,
however, and does not reduce the capacity to communicate
information. The effects of dispersion \textit{could} be taken
into account by a suitable linear transformation of the received
signal during processing. Usually, however, dispersion is accounted for by
introducing dispersion compensating elements into the transmission
link; Spans of standard, dispersive fiber are interleaved with
spans of dispersion compensating fiber in which the sign of
dispersion has been reversed, so that more slowly moving frequency
components catch up with the faster moving components.

As light propagates in an optical fiber, some of it is scattered out of the
fiber by Raman scattering leading to a loss of signal power. This is
compensated for by the periodic insertion of short spans of lasing fiber. As
the light passes through the lasing fiber, it is amplified by stimulated
emission. It is impossible to avoid a certain amount of spontaneous emission
in this process. This is amplified alongside the signal and provides a
source of additive noise called amplified spontaneous emission (ASE) noise.

Although we take full account of the effects of dispersion,
dispersion compensation and ASE noise in our analysis, these are
not the main focus this of work. A system with only these effects is
a linear system with additive noise. This is precisely the type of
system understood by Shannon many years ago\cite{Shannon48}. The additional
aspects of optical communication are the non-linearities present in the
propagation of light in an optical fiber. The most important of
these non-linearities is the optical Kerr non-linearity, by which
the refractive index of the fiber depends upon the intensity of
light in the fiber\cite{Agrawal95}.

The Kerr non-linearity causes scattering between different
frequency components of light in the fiber. Two incoming
frequency components scatter into two outgoing frequency
components. In WDM systems the effects of the Kerr non-linearity
are grouped according to the origin of the frequency components
which scatter off one another. When the incoming and outgoing
frequency components all lie in the same sub-band, the effect is
known as self-phase modulation (SPM). It is this non-linearity that
 is responsible for the possibility of soliton propagation
within a sub-band. The combination of SPM and ASE noise leads to
several mechanisms of signal distortion: fluctuations in the
amplitude and arrival times of soliton pulses, known as
Gordon-Haus jitter\cite{Gordon86}; non-linear amplification of
the phase noise\cite{Gordon90}; and, depending upon the sign
of dispersion, a non-linear instability of amplitude fluctuations
about a constant signal called the modulation
instability\cite{Hui97}. When the incoming
frequency components are in different sub-bands and scatter back
into their original sub-bands, the effect is called cross-phase
modulation (XPM). This is the effect that we will concentrate upon
here. We will show explicitly that this leads to
multiplicative noise- to a spatially and temporally random
potential due to the signal in the other sub-bands. We will find a
super-diffusive spreading of the pulse shape and a super-diffusive
spreading of arrival times. Finally, the effect of the Kerr
non-linearity when all of the incoming and 
scattered frequency components lie in different sub-bands is
known as four wave mixing (FWM). From the point of
view of the signal in any particular sub-band, it has an identical
effect to additive amplifier noise\cite{Wegener02}.

A complete WDM optical communications system consists of a series
of spans of standard, dispersive fiber interleaved with dispersion
compensating fibers and loss compensating amplifiers. The precise
ordering of these elements is known as the dispersion map. The
performance of the system can depend quite sensitively upon this
dispersion map. However, for concreteness, we will restrict
ourselves to one particular map, shown in Fig1.

Our approach will be to calculate Feynman path integrals
describing the propagation of light in a single sub-band, taking
account of scattering from signals in the other sub-bands through
XPM. The Green's functions for this propagation must be averaged
over realizations of the signal in the other sub-bands. The Green's
functions for propagation in the optical fiber turn out to be the
Green's functions of a Schr\"odinger equation with spatially and
temporally random potential. The main
calculations in this paper are the evaluation of the single and
two particle Green's functions of this Schr\"odinger equation. This
calculation is similar to calculations carried out in two other
contexts: Turbulent flow or the growth of interfaces may both be
described by the KPZ equation\cite{Kardar85,Kardar86,Kardar87}. This
may be mapped to a diffusion equation with spatially and
temporally random potential. Calculating the quadratic statistics
for these processes and averaging over the turbulent flow or noise
in the growth process is very similar to calculating the two
particle Green's function in our problem. In particular, short-range
temporal correlations allow the calculation to be reduced to
an effective single body problem for which the Feynman path
integral may be readily calculated. We will find the same
structure in our solution to the present problem. The second
problem to which our work bears similarity is the calculation of
Green's function in the presence of a quenched random potential\cite{Efetov}.
In this context, it is usual to truncate the diagrammatic series
for the single and two particle Green's functions to the Born and
Ladder series, respectively. Here we will find that this
truncation is justified by the short range temporal correlations
of the random potential. 

Our approach contrasts with other methods of including the effects of
XPM in WDM systems\cite{Mollenauer91},
in which individual scattering events between soliton pulses in different
sub-bands are considered. Our results are very similar to those
obtained for Gordon-Haus jitter\cite{Gordon86}, although of a
different physical origin.

Before going on to a detailed calculation, let us first state our
results and the physics underlying them. As a signal pulse
propagates along a particular sub-band it passes through regions
of higher and lower intensity of the total signal in the other
sub-bands. This causes the pulse to speed up and slow down due to
the Kerr non-linearity. Since the signals in the other sub-bands
are unknown to the users of any particular sub-band, this speeding
up and slowing down of the pulse has a statistical uncertainty.
The result is a diffusive spreading of the phase velocity and a
corresponding super-diffusive spreading of the distribution of pulse arrival
time; $ \langle \delta T^{2}\rangle \propto L^{3}$(to
see this requires integrating the diffusively spreading phase
velocity over the length of propagation). This is the main result
of our calculation. The functional form is very similar to that
obtained previously for Gordon-Haus jitter\cite{Gordon86}, although
its physical origin is slightly different; it is due to XPM rather than a
combination of SPM and ASE noise. In addition to the uncertainty in
the pulse arrival time, the pulse shape itself is distorted due to
the effect of XPM; since they travel at different speeds, the
different frequency components of the pulse sample slightly
different portions of the signal in the other channels and arrive
at slightly different times. Therefore, the pulse width itself
will show a super-diffusive spreading due to the effects of XPM.
The effect of XPM upon the pulse power is much weaker than its effect
upon pulse timing. The processes leading to fluctuations in pulse
power are of higher order in the (weak) non-linearity. The timing
errors induced by XPM cause the pulse power to spread over a larger
interval leading to the possibility of overlap between adjacent
pulses. This intersymbol interference allows timing errors to be
converted into amplitude errors. Numerical simulation, however, shows
that the dominant contribution to amplitude errors comes from
SPM\cite{Wegener02}, the effects of which are not studied here.

\begin{figure}
\centering
\epsfig{file=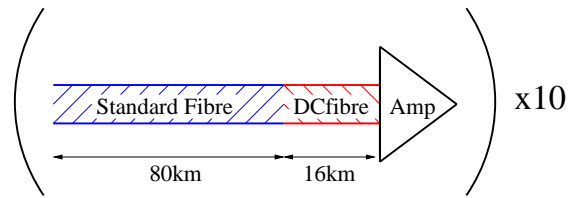,height=1in}
\caption{Schematic Diagram of Optical Fiber System}
\end{figure}

\section{The Model}

The propagation of light in an optical fiber is described by
Maxwell's equations with a dielectric constant dependent upon the
energy density of the electromagnetic field. The optical fibers
used in communications systems support only one transverse mode.
Expanding the Maxwell's equation for the low
frequency envelope of the electric field and taking the long
wavelength limit, one obtains a non-linear Schr\"odinger equation
with the roles of space and time interchanged\cite{Agrawal95}:
\begin{equation}
i\partial _{z}E(z,t) 
= 
\left(
\beta \partial _{t}^{2}
-i \alpha
\right)
E(z,t)
-\gamma \left|E(z,t)\right| ^{2}E(z,t).  
\label{NLS1}
\end{equation}
The term proportional to $\alpha$ on the right-hand side of this
equation  describes the loss of optical power from the channel and the
term proportional to $\gamma$ encodes the effects of the Kerr non-linearity.
Dividing the electric field into its components in each sub-band of the WDM
system, the low frequency envelope of the electric field in the $i^{th}$
sub-band, $E_{i}(z,t)$, is given by
\begin{eqnarray}
i\partial _{z}E_{i}(z,t) 
&=& 
\left(
\beta \partial_{t}^{2}
-i \alpha \right)E_{i}(z,t)
+V(z,t)E_{i}(z,t),  
\nonumber \\
V(z,t) 
&=& 
-2\gamma \sum_{j\neq i}|E_{j}(z,t)|^{2}.  
\label{NLS2}
\end{eqnarray}
In writing down this expression we have neglected self-phase
modulation, 
$ \gamma \left| E_{i}(z,t)\right| ^{2}E_{i}(z,t),$ 
and four-wave mixing, 
$ \gamma \sum_{j,k\neq i}E_{j}^{\ast }(z,t)E_{k}(z,t)E_{i}(z,t)$. 
This neglect of self-phase modulation and four-wave mixing may be
justified on two grounds. Firstly, there are regimes where numerical
simulation confirms this to be a good
approximation\cite{Wegener02}. Studying the system in the absence of
the effects of self-phase modulation and four-wave mixing allows us to
deduce the effects of XPM. Our aim is to determine what constraints
are imposed upon communication by these effects. In a real system,
other sources of signal distortion will impose additional constraints,
but these are not of direct concern in this paper.
Equation(\ref{NLS2}) corresponds to a
non-linear  Schr\"odinger equation with a spatially and temporally
random potential due to non-linear interaction with the signals in
the other sub-bands. The effects of loss in this system may be
accounted for by using rescaled electrical fields that maintain their
amplitude during propagation; $E_{eff,i}(z,t)=e^{\alpha z} E_i(z,t)$.
The propagation of these rescaled fields is described by
Eq.(\ref{NLS2}) with an exponential rescaling of the
strength of non-linearity,
$\gamma_{eff}=\gamma (1-e^{-\alpha L})/\alpha L$,
where $L$ is the length of a single span of fiber, and without the
loss term proportional to $\alpha$. Henceforth, we will assume the use
of these rescaled fields and drop the subscript $eff$ for brevity.

We follow Ref.\cite{Mitra01} and model the
stochastic potential by a Gaussian distribution with short-ranged
correlations in space and time; the signals in the separate
sub-bands are statistically independent and have short-ranged
temporal correlations on timescales of order $1/$bandwidth.
Relative dispersion of signals in different sub-bands leads to
short-ranged spatial correlations, which may be approximated by a
delta-function if the dispersion is sufficiently large. The
strength of the potential, 
$
\eta 
=
\int dz\langle \delta
V(z,t)\delta V(0,t)\rangle 
=
\left[ (\gamma P)^{2}/(2\beta \Delta \delta W)
\right]
\ln \left[ n_{c}/2\right]  
$, 
is obtained by summing the contributions from $n_{c}$
channels with separation $\delta W$, bandwidth $\Delta $ and
signal power, $P$. The logarithmic dependence upon the number of
channels is due to the suppression of the effects of widely
separated channels by dispersion\cite{Mitra01}. In the following
section, we will often write 
$
\langle \delta V(z,t) \delta V(0,0) \rangle
=
2 \pi \eta/ \Delta \delta(t) \delta(z)$, with an explicit
delta-function in time. This simplifies our calculations and
elucidates which contributions are most important for propagation in a
potential with short temporal correlations. It is important to
realize, however, that the potential and signal are both band-limited,
so that the correlations are not strictly delta-functional. We account
for this by regularizing the delta-function of zero argument to 
$\delta(t=0)=\Delta/2\pi$.

The system considered here consists of 10$\times$ [80km standard fiber
followed by 16km dispersion compensating (d.c.) fiber and a loss
compensating amplifier]. The input signal has power, bandwidth and channel
separation $P=5$mW, $\Delta=10$GHz and $\delta W=15$GHz, respectively. The
standard/dc fiber loss, non-linearity and dispersion parameters are 
$\alpha=0.048/0.115 \hbox{km}^{-1}$, 
$\gamma=1.2/5.1 \hbox{W}^{-1} \hbox{km}^{-1}$ 
and 
$\beta=11.0/-66.9 \hbox{ps}^2 \hbox{km}^{-1}$. 
Comparing $\gamma P/\alpha$ for the standard and d.c. fibers gives a
measure of their relative strengths of non-linearity. The power
entering the d.c fiber is a factor of  $\sim 50$ lower than that
entering the standard fiber. This, combined with the greater loss in
the d.c. fiber, leads to an effective strength of non-linearity $\sim
150$ times greater in the standard fiber. We ignore non-linearity in 
the d.c fiber in our analytical work. It may be included
in precisely the same way as in the standard fiber and does not
lead to any qualitative change in our results. Three important
lengthscales may be determined from these system parameters;
the total fiber length, 
$L_{tot}=10 \times 80km$, 
the non-linear length, 
$\eta^{-1}=223km$, 
and the dispersion length, 
$L_D=(\beta \Delta^2)^{-1}=747km $. 
The response of the system is completely determined by the ratios of
these three lengthscales. 

\section{Input Output Statistics}

The output electric field, $E_{out}(t)$, after propagating along a single
span of standard fiber, may be expressed in terms of the input electric
field $E_{in}(t)$ as
\begin{equation}
E_{out}(t)
=
\int dt^{\prime }
{\mathcal G}(t,t^{\prime })
E_{in}(t^{\prime})
+n(t),
\label{Response}
\end{equation}
where $n(t)$ is additive Gaussian white noise, with variance $N$,
representing ASE noise. ${\mathcal G}(t,t^{\prime })$ is the Green's
function of Eq.(\ref{NLS2}) for propagation from $z=0$ at time $t$ to
$z=L$ at time $t'$, with a fixed realization of $V(z,t)$. More
generally, we will use the notation ${\mathcal G}(z_1,z_2;t_1,t_2)$ to
indicate the Green's function for propagation from $t=t_1$, $z=z_1$ to
$t=t_2$, $z=z_2$. Using this notation,
${\mathcal G}(t_1,t_2) \equiv {\mathcal G}(0,L;t_1,t_2)$.
Suppression of one of the spatial indices implies propagation from $z=0$; 
${\mathcal G}(z;t_1,t_2) \equiv {\mathcal G}(0,z;t_1,t_2)$. Similarly,
${\mathcal G}(z,t) \equiv {\mathcal G}(0,z;0,t)$. It is particularly
important to note that ${\cal G}(t_1,t_2) \ne {\cal G}(t_1-t_2)$ as
would be the case for a static potential.

Dispersion
compensation is included by replacing 
${\mathcal G}(t,t^{\prime })$ 
in Eq.(\ref{Response}) with a dispersion compensated Green's function, 
${\mathcal G}_{d}(t,t^{\prime })$. This is obtained by convolving
the Green's function for the standard fiber,
${\mathcal G}(t,t^{\prime })$, with the Green's
function for the dispersion compensating fiber,
${\mathcal G}_{comp}(t,t^{\prime })$; 
${\mathcal G}_{d}(t,t^{\prime })
=
\int dt^{\prime \prime}
{\mathcal G}_{comp}(t,t^{\prime \prime })
{\mathcal G}(t^{\prime \prime },t^{\prime })$.
${\mathcal G}_{comp}(t,t^{\prime })$, has the opposite sign
of $\beta $ to ${\mathcal G}(t,t^{\prime })$ and describes propagation
with no potential, $V$. 

Propagation along a series of interleaved standard
and d.c spans is modeled by convolving a string of dispersion
compensated Green's functions. This convolution is independent of
the ordering of the Green's functions and the effects of XPM are,
therefore, independent of the dispersion map. Real systems are
sensitively dependent upon the choice of map due to the effects
of SPM, which is ignored here. Finally, due to the unitarity of 
${\mathcal G}_{d}(t,t^{\prime })$, 
adding ASE noise at each amplifier is equivalent to
adding Gaussian white noise at the receiver. Analytical calculations may,
therefore, be performed using a simplified dispersion map consisting
of $N_{s}$ standard spans followed by $N_{s}$ dispersion compensating spans and
adding Gaussian white noise at the receiver.

We now use this simple model of the system to calculate the
effects of XPM upon the input/output statistics. The quadratic
statistics are given by
\begin{eqnarray}
\langle E^*_{in}(t) E_{in}(t^{\prime}) \rangle
&=&
P \delta(t-t^{\prime}),
\nonumber \\
\langle E^*_{in}(t) E_{out}(t^{\prime}) \rangle
&=&
P \langle
{\mathcal G}_d(t,t^{\prime}) \rangle,
\nonumber \\
\langle E^*_{out}(t) E_{out}(t^{\prime}) \rangle
&=&
(P+N)
\delta(t-t^{\prime}),
\label{quadratic_stats}
\end{eqnarray}
and the quartic statistics are given by
\begin{eqnarray}
\langle \delta|E_{in}(t)|^2 \delta|E_{in}(t^{\prime})|^2 \rangle
&=&
\Delta
P^2 \delta(t-t^{\prime}) /2 \pi ,
\nonumber \\
\langle \delta|E_{in}(t)|^2 \delta|E_{out}(t^{\prime})|^2 \rangle
&=&
P^2
{\mathcal G}_d^{II}(t-t^{\prime}),
\nonumber \\
\langle \delta|E_{out}(t)|^2 \delta|E_{out}(t^{\prime})|^2 \rangle
&=& \Delta (P+N)^2 \delta(t-t^{\prime}) / 2\pi.
\label{quartic_stats}
\end{eqnarray}
The angular brackets indicate averages over the signal in the
sub-band of interest, the signals in the other sub-bands and the
amplifier noise.  $\langle {\mathcal G}_d(t,t^{\prime}) \rangle$ is
the average of the dispersion compensated, single particle Green's
function over the potential (or realizations of the signal in the
other sub-bands) and  ${\mathcal G}_d^{II}(t-t^{\prime})=\langle
|{\mathcal G}_d(t,t^{\prime})|^2 \rangle$ is the average of the
two-particle Green's function over the random potential. The input
signal is assumed to have a Gaussian distribution with power $P$
and the amplifier noise to have a Gaussian distribution with power
$N$. We have used the unitarity of the single point Green's
function for a particular realization of the potential in deriving
Eqs.(\ref{quadratic_stats}) and (\ref{quartic_stats});
${\mathcal G}(t,t)=1$. The delta-function, $\delta
(t-t^{\prime}=0)$, has been regularized as discussed in Sec.II,
to allow for the fact that the sub-band has a bandwidth $\Delta$;
$\delta(t-t^{\prime}=0)=\Delta/2 \pi$.
The averaged Green's functions, $\langle {\mathcal G}_d(t,t^{\prime})
\rangle$ and ${\mathcal G}_d^{II}(t-t^{\prime})$ are invariant under
temporal translations and are, therefore, functions of $t-t'$.

The statistics embodied by Eqs.(\ref{quadratic_stats}) and
(\ref{quartic_stats}) are non-Gaussian, since
${\mathcal G}_d^{II}(t-t^{\prime})\ne |\langle
|{\mathcal G}_d(t,t^{\prime})\rangle |^2$ in general. The
corrections to Gaussian statistics are given by the vertex
corrections to the two-particle Green's function. Alternatively,
one may say that linear propagation in the presence of a random
potential leads to non-linear propagation on average. Calculation
of the quadratic and quartic statistics in terms of Green's
functions allows a simple characterization of this non-linearity.
In the following section, we will give an explicit calculation of
the averaged Green's functions and follow this in the subsequent
sections by a discussion of their consequences for optical
communication. The quadratic statistics will be useful for our
discussion of coherent communication and the quartic statistics will
be useful for our discussion of incoherent communication.


\section{Calculations}


The average Green's functions $\langle {\mathcal G} \rangle$ and
${\mathcal G}^{II}$ may be calculated from their Feynman path
integral representations\cite{Feynman}. If we assume that the
stochastic potential is delta-function correlated in time, these
path integrals may be evaluated in closed form. These results
correspond to re-summing the Born and ladder series\cite{Efetov} in the
diagrammatic expansions of $\langle {\mathcal G} \rangle$ and
${\mathcal G}^{II}$, respectively. These series may be
used to approximate the effects of a non-delta-function correlated
potential. The diagrams that are neglected in this approximation are
small for a potential with short ranged temporal correlations; the use
of a delta-function correlated potential is a calculational tool that
emphasizes the range of validity of this approximation.
In the following , we will calculate equations of motion for $\langle
{\mathcal G} \rangle$ and ${\mathcal G}^{II}$ from their path integrals,
under the assumption that $V$ is delta-function correlated in time.
We will show how the solutions of these equations correspond to the re-summed
perturbative, ladder and Born series. Finally, the average Green's
functions $\langle {\mathcal G} \rangle$ and ${\mathcal G}^{II}$ will be
calculated allowing for the band limit of the sub-band and potential
due to the other sub-bands.

\subsection{Single particle Green's function}

Recalling that the roles of space and time have been interchanged here
compared with their more familiar roles in quantum mechanics, the
single particle Green's function is written in the form
\begin{eqnarray}
\langle {\mathcal G} (L,t) \rangle 
&=& 
\langle 
\int_{t(0)=0}^{t(L)=t} Dt(z)
\exp \left[ 
-\int_0^L dz \frac{i}{2 \beta} \left( \partial_z t(z) \right)^2
\right.
\nonumber\\
& &
\left.
-i\int_0^L dz V(z,t(z)) \right] \rangle  
\nonumber \\
&=& 
\int_{t(0)=0}^{t(L)=t} Dt(z) 
\exp \left[ -\int_0^L dz \frac{i}{2 \beta}
\left( \partial_z t(z) \right)^2 \right.  
\nonumber \\
& & 
\left. -\frac{1}{2} \int_0^L \int_0^L dz dz^{\prime}
\langle 
V(z,t(z)) V(z^{\prime},t(z^{\prime})) 
\rangle \right],  
\label{Path_integral_for_G}
\end{eqnarray}
where the angular brackets denote averages over realizations of $V$. We have
carried out the average over $V$ in order to obtain the second line,
assuming that $V$ is a Gaussian random variable.

Taking the derivative of Eq.(\ref{Path_integral_for_G}) with respect
to $L$\cite{Feynman}, we obtain the following equation of motion for 
$\langle {\mathcal G} \rangle$:
\begin{eqnarray}
\lefteqn{
\left( 
-i \frac{\beta}{2} \partial_t^2 -\partial_L 
\right) 
\langle {\mathcal G}(L,t) \rangle }
\nonumber\\
&=&
\delta(L) \delta(t)  
+ \int_0^L \int_{-\infty}^{\infty} dt^{\prime}dz 
\langle V(L,t)V(z,t^{\prime}) \rangle
\nonumber\\
& &
\times
\langle {\mathcal G}(L-z,t-t^{\prime}) \rangle 
\langle {\mathcal G}(z,t^{\prime}) \rangle.  
\label{G_Equation_of_motion}
\end{eqnarray}
This equation is local in time for a stochastic potential that is
delta-function correlated in time; 
$\langle V(z,t) V(0,0) \rangle =(2 \pi \eta/\Delta) \delta(t) \delta(z)$. 
The solution of this local equation of motion is 
$\langle {\mathcal G}(L,t) \rangle= e^{-\eta L/2} {\mathcal G}_0(L,t) $, 
where ${\mathcal G}_0(L,t)$ is the Green's function of the
operator $\left( -i \beta \partial_t^2/2 -\partial_L \right)$,
\textit{i.e.} the Green's function for propagation in the absence of $V$.
This result was first obtained in Ref.\cite{Mitra01}.
Equation(\ref{G_Equation_of_motion}) may also be
obtained diagrammatically by re-summing the Born series; it is nothing
but the Dyson's equation for $\langle {\mathcal G}(L,t) \rangle$
obtained by re-summing this series\cite{Efetov}.
The temporal delta-function correlation of $V$ implies that the propagators of
$V$ must not cross, thus restricting the contributing diagrams to the Born
series. Figure 2. shows a symbolic representation of 
Eq.(\ref{G_Equation_of_motion}) after multiplying on the left by
${\mathcal G}_0$. The dashed line indicates propagators of $V$, the
solid line indicates ${\mathcal G}_0$ and the thick solid line
indicates 
$\langle {\mathcal G} \rangle$. This diagram is to be interpreted in
real space or momentum space in the usual way. The momentum/frequency
space solution is 
\begin{equation}
\langle {\mathcal G} (p,\omega) \rangle
=
i \left(
p-\beta \omega^2 +i \eta/2
\right)^{-1},
\label{G_momentum_frequency}
\end{equation}
where $p$ is the wavevector conjugate to $z$ and $\omega$ is the
angular frequency conjugate to $t$.
For later calculations, it is useful to have $\langle {\mathcal G}
\rangle$ in a mixed, position/frequency notation, where it is given by
\begin{equation}
\langle {\mathcal G} (z,\omega) \rangle
=
\exp \left[
-\left(
\eta/2-i \beta \omega^2 \right) z
\right].
\label{G_position_frequency}
\end{equation}
\begin{figure}
\epsfig{file=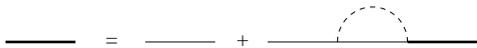,width=2.5in}
\caption{Born series for $\langle {\cal G} \rangle$.}
\end{figure}
Dispersion compensation is achieved by temporally convolving ${\cal
  G}(L,t)$ for the standard fiber with that of the compensating 
fiber (or alternatively multiplying together the position/frequency space
Green's functions). Ignoring non-linearity in the compensating fiber,
we find
${\mathcal G}_{comp}(L,t)={\mathcal G}_0^*(L,t)$. The dispersion
compensated Green's function is then given by
\begin{equation}
\langle {\mathcal G}_d \rangle(L,t)=e^{-\eta L/2} \delta (t).
\end{equation}
On average, therefore, the effects of XPM non-linearity  separate from those
of dispersion. Moreover, the effect of XPM is to dephase the
propagating electric field.

\subsection{Calculation of ${\mathcal G}^{II}$}

The calculation of ${\mathcal G}^{II}$ from its path integral proceeds in a
very similar manner to the calculation of $\langle {\mathcal G} \rangle$.
After writing down the path integral representation of ${\mathcal G}^{II}$,
we perform the average over $V$ and deduce an equation of motion for 
${\mathcal G}^{II}$ by differentiating with respect to $L$. This differential
equation is local in time when the correlation function of $V$ is a temporal
delta-function. The equation of motion is most easily solved in the
frequency domain, where it is recognizable as the re-summed ladder series. As
before, the set of contributing diagrams is restricted due to the
requirement that the propagators of $V$ may not cross. We will first of all
derive an expression for ${\mathcal G}^{II}(\omega,q)$ in the absence of
dispersion compensation from its path integral representation. We will
verify that this result is identical to that obtained by summing the ladder
series of diagrams. In fact, closed expressions for ${\mathcal G}^{II}(L,t)$
are most readily obtained by re-summing this series directly in the
length/frequency domain. We explicitly carry out this re-summation to obtain
expressions for ${\mathcal G}^{II}(L,\omega)$ and its dispersion compensated
counterpart ${\mathcal G}^{II}_d(L,\omega)$.

We will first derive a path integral and equation of motion for the Green's
function, 
${\mathcal G}^{II}(L;\tau_1,\tau_2)
=\langle {\mathcal G}(L,\tau_1) {\mathcal G}^*(L,\tau_2) \rangle$. 
In the end we will be interested in 
${\mathcal G}^{II}(L,t)={\mathcal G}^{II}(L;\tau_1=t,\tau_2=t)$. The path
integral representation of ${\mathcal G}^{II}$ is
\begin{widetext}
\begin{eqnarray}
{\mathcal G}^{II}(L;\tau_1,\tau_2)  
&=& 
\langle {\mathcal G}(L,\tau_1) {\mathcal G}^*(L,\tau_2) \rangle  
\nonumber\\
&=& 
\langle 
\int_{\tau_1(0)=0,\tau_2(0)=0}^{\tau_1(L)=\tau_1,\tau_2(L)=\tau_2} 
D\tau_1(z) D \tau_2(z) 
\exp 
\left[
\int dz 
\left( 
i \frac{(\partial_z \tau_1)^2}{2 \beta} 
-i \frac{(\partial_z \tau_2)^2}{2 \beta}
\right) 
\right]  
\nonumber \\
& & 
\times 
\exp
\left[ 
-\frac{1}{2}\int dz 
\left( 
-i V\left( z, \tau_1(z)\right) 
+i V\left( z, \tau_2(z) \right) 
\right) 
\right]  
\rangle
\nonumber \\
&=& 
\int_{\tau_1(0)=0,\tau_2(0)=0}^{\tau_1(L)=\tau_1, \tau_2(L)=\tau_2}
D\tau_1(z) D \tau_2(z) 
\exp 
\left[
\int dz 
\left( 
i \frac{(\partial_z\tau_1)^2}{2 \beta} 
-i \frac{(\partial_z \tau_2)^2}{2 \beta} 
\right) 
\right]
\nonumber \\
& & 
\times 
\exp 
\left[
-\frac{1}{2}\int dz dz^{\prime}
\langle 
V\left( z,\tau_1(z) \right) 
V\left( z^{\prime},\tau_1(z^{\prime})\right) 
\rangle 
\right]  
\nonumber \\
& & 
\times 
\exp \left[ 
-\frac{1}{2}\int dz dz^{\prime}
\langle 
V\left(z, \tau_2(z) \right) 
V\left(z^{\prime},\tau_2(z^{\prime})\right) 
\rangle 
\right]  
\nonumber \\
& & 
\times 
\exp \left[ 
\int dz dz^{\prime}
\langle 
V\left( z,\tau_1(z)\right) 
V\left( z^{\prime},\tau_2(z^{\prime})\right) 
\rangle 
\right]
\label{Path_integral_for_GII}
\end{eqnarray}
\end{widetext}
The average over $V$ has been carried out in passing from the penultimate to
final line of Eq.(\ref{Path_integral_for_GII}). The equation of motion
for ${\mathcal G}^{II}$ is derived by differentiating both sides of
this equation with respect to $L$. The resulting differential equation is
local in time when $V$ has delta-function correlations in time. It is
slightly easier to make the assumption of temporally delta-function
correlations for $V$ in Eq.(\ref{Path_integral_for_GII}) before
differentiating. The path integral then reduces to
\begin{eqnarray}
{\cal G}^{II}(L;\tau_1,\tau_2) 
&=&
\int_{\tau_1(0)=0,\tau_2(0)=0}^{\tau_1(L)=\tau_1, \tau_2(L)=\tau_2}
D\tau_1(z) D \tau_2(z) 
\nonumber\\
& & 
\times 
\exp 
\left[\int dz 
\left( 
i \frac{(\partial_z \tau_1)^2}{2 \beta}
-i \frac{(\partial_z \tau_2)^2}{2 \beta} 
+\eta 
\right.\right.
\nonumber\\
& &
\left.
\left.
- \eta \frac{2 \pi}{\Delta}
\delta \left( \tau_1(z)-\tau_2(z) \right) 
\right) 
\right].
\label{Path_integral_for_GII2}
\end{eqnarray}
This result is very similar to that obtained in the analysis of the
diffusion equation with a spatially and temporally random 
potential\cite{Kardar85}.
In that case, the path integral for the square of the wavefunction reduced
to that of two interacting random walks. Here the path integral for 
${\mathcal G}^{II}$ reduces to that of two interacting Schr\"odinger particles,
or waves, propagating in opposite directions. Differentiating 
Eq.(\ref{Path_integral_for_GII2}) with respect to $L$, we obtain the
following local equation of motion for ${\mathcal G}^{II}$.
\begin{eqnarray}
& & 
\left( 
\partial_L 
-i\frac{\beta}{2}(\partial_{\tau_1}^2 -\partial_{\tau_2}^2 ) 
+\eta 
\right) 
{\mathcal G}^{II}(L;\tau_1,\tau_2)
\nonumber \\
&=& 
\delta(L) \delta (\tau_1) \delta (\tau_2) 
+\eta \frac{2 \pi}{\Delta}
\delta (\tau_1-\tau_2) 
{\mathcal G}^{II}(L;\tau_1,\tau_2).
\label{GII_Equation_of_motion}
\end{eqnarray}
This equation is most easily solved by first changing coordinates to the
mean time, $t=(\tau_1+\tau_2)/2$, and time difference,
$T=(\tau_1-\tau_2)/2$. After Fourier transforming in space and time we find
\begin{eqnarray}
-i
\left( q - \frac{\beta}{2} \omega \Omega + i\eta \delta(0) \right)
{\mathcal G}^{II}(q;\omega,\Omega) 
\nonumber\\
= 
1
+ 
\frac{2 \pi \eta}{\Delta} 
\int \frac{d \Omega}{2 \pi} 
{\mathcal G}^{II}(q;\omega,\Omega),  
\label{Dyson}
\end{eqnarray}
where $\omega$ and $\Omega$ are the angular frequencies conjugate to $t$ and
$T$, respectively and $q$ is the wave-vector conjugate to
$L$. Equation(\ref{Dyson}) is equivalent to the ladder series shown in
Fig.2. In order to make this 
correspondence clear, the effects of non-linearity have been divided into
two parts; the term proportional to $\eta$ on the left hand side of Eq.(\ref%
{Dyson}) takes account of self-energy correction to the single particle
Green's function and the term proportional to $\eta$ on the right hand side
takes account of vertex corrections. Following the usual convention,
we denote the solution of Eq.(\ref{Dyson}) in the absence of vertex
corrections by 
$\Pi(q;\omega,\Omega)=-i\left(q-\beta\omega \Omega/2 + i\eta
\right)^{-1}$. 
One may check by direct substitution of 
$\langle {\mathcal G}(q,\omega) \rangle $
from Eq.(\ref{G_momentum_frequency}) that 
$\Pi(q;\omega,\Omega)
=
\int d p/2 \pi \langle {\mathcal G}(q+p,\omega+\Omega)
\rangle \langle {\mathcal G}^*(p,\Omega) \rangle$. 
Finally, using the notation 
$\Pi(q,\omega)=\int d \Omega/ 2 \pi \Pi(q;\omega,\Omega)$, the
solution of Eq.(\ref{Dyson}) is
\begin{eqnarray}
{\mathcal G}^{II}(q,\omega) 
&=& 
\frac{\Pi(q,\omega)}{1- \frac{2 \pi \eta}{\Delta} 
\Pi(q,\omega)}  
\nonumber \\
&=& 
\Pi(q,\omega) 
\left( 1 + \frac{2 \pi \eta}{\Delta} \Pi(q,\omega)
\right.
\nonumber\\
& & 
\left.
+ 
\left(\frac{2 \pi \eta}{\Delta} \right)^2 \Pi(q,\omega)^2 +... \right),
\end{eqnarray}
which corresponds to the ladder diagram shown in Fig.3.
\begin{figure}
\epsfig{file=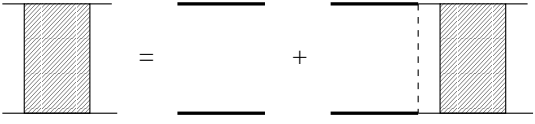,width=2.8in}
\caption{Ladder series for ${\cal G}^{II}$.}
\end{figure}

The modification to these results due to dispersion compensation is not
quite as simple as the modification to $\langle {\mathcal G} \rangle$.
Although it is possible to derive a path integral expression for 
${\mathcal G}^{II}_d$ akin to Eq.(\ref{GII_Equation_of_motion}), it is
easiest to derive the dispersion compensated
${\mathcal G}^{II}$ using diagrammatic perturbation theory in the
length/frequency domain.  The appropriate diagrammatic series is shown
in Fig.4. The result of summing this series is
\begin{eqnarray}
& &
{\mathcal G}^{II}_d (L,\omega) 
\nonumber\\
&=& 
\Pi_d(L, \omega) 
\nonumber \\
& & 
+ 
\frac{2 \pi \eta}{\Delta} 
\int_0^L dx 
\Pi (x,\omega) 
\Pi_d (L-x,\omega)  
\nonumber \\
& & 
+ \left(\frac{2 \pi \eta}{\Delta} \right)^2 
\int_0^L dx \int_0^x dy
\Pi(x,\omega) \Pi (y,\omega) \Pi_d (L
-y,\omega) 
\nonumber\\
& &
+... 
\label{GIId_diagrams}
\end{eqnarray}
$\Pi(z,\omega)$ is given by
\begin{eqnarray}
\Pi(z,\omega)
&=&
\int \frac{d \omega^{\prime}}{2 \pi} 
\langle {\mathcal G}(z, \omega^{\prime}+\omega/2) \rangle
\langle {\mathcal G}^* (z, \omega^{\prime}-\omega/2) \rangle
\nonumber \\
&=& 
e^{-\eta z} 
\int \frac{d \omega^{\prime}}{2 \pi}
e^{i \beta
\omega^{\prime}\omega z}
\label{P}
\end{eqnarray}
and $\Pi_d(z,\omega)$ is the dispersion compensated analogue of 
$\Pi(z,\omega)$, given by
\begin{eqnarray}
\Pi_d(z,\omega) 
&=& 
\int \frac{d \omega^{\prime}}{2 \pi} 
\langle {\mathcal G}(z, \omega^{\prime}+\omega/2) \rangle
{\mathcal G}_0^* (L,\omega^{\prime}+\omega/2)
\nonumber \\
& &
 \;\;\;\;\;\;\ 
\times 
\langle {\mathcal G}^* (z, \omega^{\prime}-\omega/2) \rangle
{\mathcal G}_0 (L,\omega^{\prime}-\omega/2)  
\nonumber \\
&=& 
e^{-\eta z} \int \frac{d \omega^{\prime}}{2 \pi} 
e^{i \beta \omega^{\prime}\omega(z-L)}.  
\label{Pid}
\end{eqnarray} 
The calculation of ${\mathcal G}^{II}$ proceeds by substituting
$\Pi_d(z,\omega)$ and $\Pi(z,\omega)$ 
into Eq.(\ref{GIId_diagrams}), carrying out the frequency integrals in each
term, followed by the integrals over length, and finally summing the
contributions from each term. In carrying out this procedure, one must take
careful account of the support for the various frequency integrals. The band
limitation of the sub-band implies that the angular frequency carried by
each internal Green's function must lie between $-\pi \Delta$ and $+\pi
\Delta$. Carrying out the frequency integrals with this restriction leads to
very complicated expressions. An alternative approach, which yields much
more manageable expressions, is to take into account the fact that the
stochastic potential is not strictly delta-function correlated in time, but
is band-limited to lie between $-\pi \Delta$ and $+\pi \Delta$. We use the
fact that $V$ has short range correlations in time to justify neglecting
terms other than those of the Born and ladder series in our diagrammatic
calculation, but allow for the band limitation of $V$ in calculations of
internal frequency integrals.

Let us illustrate this by explicitly calculating the second order term in
Eq.(\ref{GIId_diagrams}). This is given by
\begin{eqnarray}
& & 
\int_0^L dx \int_0^x dy 
\Pi (x,\omega) \Pi (y,\omega) \Pi_d (L-y,\omega)  
\nonumber \\
&=& 
\int_0^L dx \int_0^x dy 
\int 
\frac{d \omega_1}{2 \pi} 
\frac{d \omega_2}{2 \pi} 
\frac{d \omega_3}{2 \pi}
\nonumber\\
& &
\times 
\exp \left[ 
i \beta \omega(\omega_3-\omega_2)x
+i \beta \omega (\omega_2- \omega_1)y \right]  
\nonumber\\
&=& 
\left( \frac{\Delta}{2 \pi} \right)^3 
e^{-\eta L} 
\int_0^{\pi \beta \omega \Delta L} 
dX \int_0^X dY 
\frac{\sin X}{X} \frac{\sin Y}{Y}  
\nonumber\\
&=& 
\left( \frac{\Delta}{2 \pi} \right)^3 
L e^{-\eta L} \frac{1}{2} \left.
\left( \frac{\hbox{Si}[X]}{X} \right) \right|_{X=\pi \beta \omega \Delta L},
\end{eqnarray}
where $Si[X]=\int_0^X du \sin u/u$. We have carried out the frequency
integrals in passing between the second, third and fourth lines. The
integration 
variables have been changed from $\omega_1$, $\omega_2$ and $\omega_3$
to $\omega_1$, $\omega_3-\omega_2$ and $\omega_2-\omega_1$. Integrations have
been carried out over the domains $-\Delta/2 \le \omega_1,
\omega_3-\omega_2,\omega_2-\omega_1\le \Delta/2$, the first corresponding to
the band-limitation of the signal in the sub-band and the second and third
to band-limitation of the stochastic potential leading to a restriction on
the transfer of frequency at each vertex.
\begin{figure}
\centering
\epsfig{file=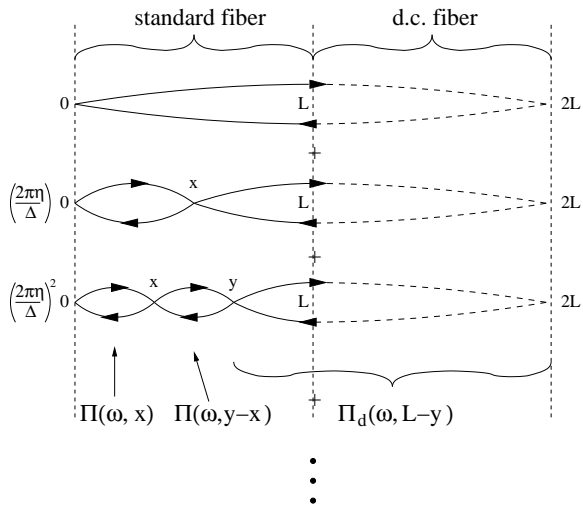,width=3in}
\caption{Position-space diagrammatic series for ${\cal G}_d^{II}(L,\omega)$.}
\end{figure}

Calculating each term in the series in this way and re-summing, we obtain our
final result
\begin{equation}
{\mathcal G}_d^{II}(L,\omega)
= 
\frac{\Delta}{2 \pi} \exp \left. 
\left[ 
-\eta L \left( 1- \frac{\hbox{Si}[X]}{X} \right) 
\right] 
\right|_{X=\pi \beta\omega \Delta L},  
\label{GII}
\end{equation}
for dispersion compensated propagation. The Green's function for
uncompensated propagation is calculated similarly replacing $\Pi_d$
with $\Pi $ in Eq.(\ref{GIId_diagrams}). The result is
\begin{equation}
{\mathcal G}^{II}(L,\omega)
= 
{\mathcal G}_d^{II}(L,\omega) \left. 
\frac{\sin X}{X} \right|_{X=\pi \beta \omega \Delta L}.
\label{GII2}
\end{equation}
The prefactor of $\sin X/X$ describes the dispersive spreading of a pulse
propagating in the absence of non-linearity. The distortion of the pulse due
to XPM is described by ${\mathcal G}_d^{II}(L,\omega)$. There are two
important limits of Eq.(\ref{GII}). At small frequencies, 
$\hbox{Si}[X]/X \rightarrow 1-X^2/18$. As a consequence, 
${\mathcal G}_d^{II}(L,\omega)$
tends to a Gaussian of width 
$\sqrt{\langle \omega^2 \rangle} = 2 \pi
\Delta_{eff} = (3 \Delta/\pi) \sqrt{ L_D^2/\eta L_{tot}^3}$. The
interpretation of this width as an effective bandwidth will be discussed
below. At large frequencies, ${\mathcal G}_d^{II}(L,\omega)$ tends to a
constant. These limits imply that ${\mathcal G}_d^{II}(L,t)$ has Gaussian
tails at large times and approaches a delta-function at short times. We will
discuss the consequences of these results for communication in the following
sections.


\section{Amplitude and Timing Jitter}

The results of sections III and IV give the input/output statistics
of the WDM optical fiber in terms of its single and two-particle Green's
functions. In this and the following section, we shall use these results to
discuss the effectct of XPM upon communication. 

Firstly, we discus two operational measures of the efficiency of an
optical communications system. Our discussion will be quite brief,
since these considerations are not central to the main message of this
paper. Nevertheless, it is important to demonstrate that analytical
results such as Eqs.~(\ref{GII}) and ~(\ref{GII2}) can be used to enhance
understanding from this operational perspective. A common way of
characterizing signal distortion in optical communications is by
amplitude and timing jitter. These measures of system performance are
specific to digital communication. A digital message is made up of a string of
pulses or marks representing 1s, and spaces representing zeros. As a pulse
propagates along a noisy channel, it speeds up and slows down in a random
way leading to a distribution of arrival times(defined as the centroid
in time of the power distribution at the output) at the receiver known as
timing jitter. Similarly, the total power of the pulse fluctuates leading to a
distribution of amplitudes of the received pulse known as amplitude
jitter. The distributions of amplitude and timing fluctuations need
not be independent\cite{Falkovich01}.
In fact, the terms amplitude and timing jitter are often
reserved for the variance in amplitude and arrival times of a
pulse. This is the sense in which we shall uses these terms from now
on. In addition to jitter, the shape of the individual pulses may be distorted
during propagation. We do not discus these distortions here.

The effects of SPM and ASE upon the amplitude and timing jitter of
soliton, or soliton-like, pulses has been considered in a number of
works, starting with Ref.\cite{Gordon86}. All
of these works expand in small fluctuations- induced by the
additive ASE noise- about a soliton solution. Such analyses lead
to the conclusion that amplitude jitter is not very significant
and that timing jitter leads to an $L^3$ dependence of the variance of arrival
times. This latter result is known as Gordon-Haus jitter after its
discovery in Ref.\cite{Gordon86}. A recent work of Falkovich
\textit{et al}\cite{Falkovich01} has used a saddle point approximation within a
Martin-Siggia-Rose field theory for this propagation to obtain a
full joint distribution of amplitude and timing errors. This work
shows in a particularly transparent way, the connection between
the method of optimal fluctuations in field theory and large
deviations in statistics. Here we use Green's function techniques to
consider the contribution of XPM to jitter. Despite its different
physical origin, the resulting jitter has a broadly similar dependence
upon system parameters to Gordon-Haus jitter.

In principle, the output for an arbitrary input pulse shape may be
calculated, accounting for the effects of XPM, in terms of Green's
functions for propagation along the fiber. The amplitude and
timing jitter may then be obtained by taking appropriate averages. 
This calculation may in practice prove rather cumbersome.
However, the functional dependence of the jitter upon system parameters is
expected to be broadly independent of the input pulse shape (the magnitude of
the jitter will show a weak dependence upon the pulse shape). For
simplicity, therefore, we present approximate expressions for 
amplitude and timing jitter due to XPM, by considering the output for
constant and delta-function inputs, respectively. The simple
expressions obtained in this way demonstrate the functional dependence
of the jitter upon the system parameters. In particular, we find that the
variance of arrival times is proportional to $L^3$, whereas
amplitude jitter is largely independent of XPM . These functional
dependences are the same as those due to Gordon-Haus jitter although
the physical origin is somewhat different\cite{Gordon86}.

Timing jitter may be found by considering the response to an input
delta-function in time. A real signal pulse is
band-limited, but it may be thought of as being made up of many delta
functions. The use of a delta-function input allows us to calculate
the functional form of the timing jitter in an analytically tractable
way. The dependence of the resulting jitter upon system parameters is
expected to be broadly similar for arbitrary input pulse shapes.
Without XPM, the output
electric field resulting from a delta function input is 
$E_{out}(t)={\mathcal G}_d(t)\propto \delta(t)$ 
and the modulus of the electric field is given by 
$|E_{out}(t)|^2=|{\mathcal G}_d(t)|^2\propto \delta(t)$. Introducing XPM
broadens this response. Different frequencies making up the pulse see a
slightly different stochastic potential and arrive at slightly different
times. Since the signals in the neighboring sub-bands are unknown, this
broadening leads to uncertainty in the arrival time. Averaging the response
to a delta function input over realizations of the signals in the other
sub-bands, therefore, we find that ${\mathcal G}_d^{II}(t)=\langle |{\mathcal G%
}_d(t)|^2 \rangle$ gives a measure of the distribution of arrival
times. Using this interpretation and the results of the preceding
section on the calculation of ${\mathcal G}_d^{II}$, we find 
$\langle t^2 \rangle=(2 \pi
\Delta_{eff})^{-2}\propto L_{tot}^{3}\eta$; XPM causes a
super-diffusive spreading of arrival times. This functional dependence
of the timing jitter upon the fiber length is identical to that found by
Gordon and Haus for the combined effects of ASE noise and SPM\cite{Gordon86}.
Notice that this contribution to timing jitter arises entirely from
vertex corrections to the 2-particle Green's function,
${\mathcal G}_d^{II}(t)$; if we ignore vertex corrections by making the
approximation ${\mathcal G}_d^{II}(t)=|\langle {\cal G}_d(t)
\rangle|^2$ and substituting Eq.~(\ref{G_position_frequency}), we find
no contribution to timing jitter.

Using Eq.(\ref{Response}) to calculate the statistics of the output for inputs
held constant at $E_{in}(t)=0$ and $E_{in}(t)=A$ gives an estimate of 
the amplitude statistics of a 1 and 0 in the output bit
stream\cite{Marcuse90}. For a zero input, the output has the same Gaussian
distribution as the noise. For an input of constant amplitude $A$,
\begin{eqnarray}
\langle E_{out}(t) \rangle_1 
&=& 
A \langle {\mathcal G}_d(\omega=0)\rangle,
\nonumber \\
\langle |E_{out}(t)|^2 \rangle_1 
&=& 
(A^2+N)\delta(0),  
\nonumber \\
\langle \delta |E_{out}|^2(t) \delta |E_{out}|^2(t) \rangle_1 
&=& 
A^4 \left[
\langle |{\cal G}_d(\omega=0)|^4\rangle-\delta(0)^2
\right]
\nonumber\\
& &
+
N(2A^2+N)\delta(0)^2 
\nonumber\\
& \approx&
N(2A^2+N)\delta(0)^2, 
\label{Amplitude_Jitter}
\end{eqnarray}
where the delta-function of zero argument is to be understood as 
$\Delta/2\pi=\delta(0)$ for our band-limited sub-band.
The evaluation of these results is outlined in the Appendix.
The output for a 1 input is not Gaussian: a Gaussian fit to 
$\langle E_{out}(t) \rangle_1$ and $\langle |E_{out}(t)|^2 \rangle_1$ does not
reproduce $\langle \delta |E_{out}|^2(t) \delta |E_{out}|^2(t)
\rangle_1$.
These results also predict that the contribution of XPM to amplitude
jitter is small. Equation~(\ref{Amplitude_Jitter}) depends upon the
evaluation of $\langle |{\cal G}(\omega=0)|^4\rangle$, which we have
not carried out here. A full evaluation of this 4-particle Green's
function would be required to give a full expression for amplitude
jitter. If, however, we approximate 
$
\langle |{\cal G}(\omega=0)|^4\rangle
\approx
\langle |{\cal G}(\omega=0)|^2\rangle^2
=\delta(0)^2
$
then amplitude jitter shows no dependence upon XPM. Corrections to
this result arise from higher order vertex corrections to the 4-particle
Green's function and are small for weak non-linearity. In particular,
these vertex corrections are of higher order than those that give rise
to timing jitter. The predominant effect of XPM, therefore, is to induce timing
errors. This conclusion is in accord with other works which
have attempted to model the effects of XPM as a modification to the
Gordon-Haus results for soliton jitter\cite{Mollenauer91}. In these
works, individual scattering events between 
soliton-like pulses in different sub-bands of the WDM system are considered
and the amplitude jitter due to scattering ignored.


\section{Information Theory}

In this section, we will discuss how the input/output statistics
calculated in sections III and IV may be used to construct
general arguments about the maximum rate of communication via a
WDM system. This is still an open problem. The only systems for
which exact answers about information capacity may be obtained are
those with Gaussian input/output statistics. We will use the
results for Gaussian systems to obtain lower bounds  upon and
approximations to the capacity of the present, non-Gaussian system.

The ideas of information theory propounded by Shannon are very
much akin to the ideas of statistical mechanics. For a signal made up of a
sequence of letters (or field values), chosen from some alphabet
(or range of values) one may define a probability distribution,
$p(x)$, for the probability that a particular letter takes the
value $x$. The amount of information per letter in such a sequence
is given by the entropy of this distribution; $H(X) \equiv
H[p(x)]=-\sum_x p(x) \log p(x)$. If the logarithm is taken in base
2, this gives the amount of information per letter in bits. One
way to understand this is to think of compressing a signal.
Compression involves removing redundancy in the signal (or
alternatively correlations between letters) and so increasing the
amount of information per letter. In this way, a signal with
higher entropy will carry a higher density of information.

When one attempts to communicate via a channel, each letter, $x$,
of the input signal will be received as some letter, $y$, at the
output. Allowing for the uncertainty in the transmission process,
the output for an input, $x$, will be given by some conditional
distribution $p(y|x)$. For an information theorist, this defines
the channel of communication. For each input letter, the output
may take a range of values. The entropy of this additional
spreading at the output is given by the conditional entropy;
$H(Y|X)=\sum_x p(x) H(Y|X=x)=-\sum_x p(x) [ \sum_y p(y|x)
\log p(y|x) ]$.

When a sequence of letters is transmitted, the amount of
information in common between the input and output, otherwise know
as their mutual information is given by the total entropy of the
output signal minus the additional entropy due to signal
distortion;
\begin{equation}
I(X;Y)=H(Y)-H(Y|X).
\label{mutual information}
\end{equation}
When using a particular channel, the conditional distribution
$p(y|X)$ is fixed and the remaining freedom is in the choice of
input distribution, $p(x)$. The idea is that if certain letters
are more distorted than others, it pays one to use them less
frequently, even though this means a reduction in the amount of
information that the input signal can carry. The maximum
transmitted information density that can be achieved is given by
functionally optimizing the mutual information over the choice of
input signal distribution;
\begin{equation}
C= \max_{p(x)} I(X;Y).
\label{Capacity}
\end{equation}
Shannon formulated these ideas rigorously and showed that this is
not only an upper bound upon the rate of transmission of
information, but that it is also an achievable bound. A summary of
these basics of information theory may be found in
Ref.\cite{Cover91}. The functional optimization in
Eq.(\ref{Capacity}) is carried out under the various system
constraints, such as the average signal power. In fact, the only
case for which there exists an explicit analytical solution is
when the joint distributions of input and output signal are
Gaussian. In this case, the capacity is given by
\begin{equation}
C
= 
\log 
\left[ 
\frac{\langle x^*x \rangle \langle y^* y \rangle}
{\left|
\begin{array}{cc}
\langle x^* x \rangle &\langle x^* y \rangle\\
\langle y^* x \rangle &\langle y^* y \rangle \end{array} 
\right|}
\right].
\end{equation}
When the signal is band-limited, the corresponding expressions for
the capacity per unit bandwidth, or the spectral efficiency, is
given by integrating over frequency;
\begin{equation}
c
= 
\frac{1}{2 \pi \Delta}
\int d \omega 
\log 
\left[ 
\frac{
\langle x_{\omega}^*x_{-\omega} \rangle
\langle y_{\omega}^* y_{-\omega} \rangle} 
{\left|
\begin{array}{cc}
\langle x_{\omega}^* x_{-\omega} \rangle 
&\langle x_{\omega}^* y_{-\omega} \rangle\\
\langle y_{\omega}^* x_{-\omega} \rangle 
&\langle y_{\omega}^*y_{-\omega}\rangle
\end{array} 
\right|} \right].
\label{Gaussian_Capacity}
\end{equation}
The factor of $1/2\pi$ arises because $\omega$ is the angular
frequency conjugate to $t$.

We are now in a position to use these results from information
theory, together with our knowledge of the quadratic and quartic
statistics of the WDM system to obtain a lower bounds and
estimates of the capacity of the WDM optical communications
system. A rigorous lower bound on the capacity of a channel
defined by a conditional distribution, $p(y|x)$, may be found as
follows: The quadratic input/output statistics are calculated for
a Gaussian input distribution. A channel whose statistics are
Gaussian and whose quadratic statistics correspond to those of the
actual channel, will have capacity given by Eq.(\ref{Gaussian_Capacity}).
This gives a lower bound on the capacity of the actual channel. A
proof of this is given in Ref.\cite{Mitra01}. The basic idea
behind it is that deviations from Gaussian statistics correspond
to extra correlations in the noise that the channel adds to the
signal. These extra correlations allow one to make 
progress in determining which part of the received signal is noise
and which part is signal. Ignoring these correlations- as one does
in the Gaussian approximation- leads one to underestimate the
extent to which noise may be removed from the signal and so to an
underestimate of the system capacity. Notice that since a Gaussian
distribution maximizes the entropy for a fixed variance, any
deviation from Gaussianity will increase correlations.

We are interested in two ways of using the WDM optical fiber
system. In the first case, both the amplitude and phase of the
electrical signal are used to communicate. We call this coherent
communication. In the second situation, only the amplitude of the
electrical field is used to communicate. This latter case is more
representative of current optical communication systems.
Coherent communication was considered by Mitra and Stark in
Ref.\cite{Mitra01}. The quadratic statistics given by
Eq.(\ref{quadratic_stats}) are combined with Eq.(\ref{Gaussian_Capacity})
to give the following expression for the spectral efficiency in
terms of the single particle Green's function of the channel:
\begin{equation}
 c
 \ge
\int \frac{d \omega}{2 \pi \Delta} 
\log \left[ 1+P_{eff}(\omega)/N_{eff}(\omega) \right]
\end{equation}
with the effective signal and noise powers given by
\begin{eqnarray}
 P_{eff}(\omega)
&=&
P| \langle {\cal G}_d(\omega) \rangle|^2
\nonumber\\
N_{eff}(\omega)
&=&
P\left(1-| \langle {\cal G}_d(\omega) \rangle|^2 \right)
+N,
\label{coherent_capacity}
\end{eqnarray}
respectively. This result is a modification of the famous
Shannon\cite{Shannon48} 
result for the capacity of a linear channel with additive white
noise, allowing for the conversion of signal power to noise power
by scattering from signals in the other sub-bands. The
single-particle Green's function calculated in section 4, $\langle
{\cal G}_d(L,\omega) \propto \exp [-{\hbox{const.}}\times P^2 L]$ is
substituted into Eq.(\ref{coherent_capacity}) to obtain the
functional dependence of the capacity upon system parameters. The result is
sketched in Fig.5. There are two main features worthy of
comment. The first is the peak in capacity for a particular power;
increasing the signal power relative to the noise power is
productive up to a point, but when the signal power is very large,
non-linear interactions between signals in the different sub-bands
become the dominant mechanism of signal distortion. The appearance
of an optimal power is a common feature of the effects of
non-linearity. It occurs in the Gordon-Haus jitter of solitons\cite{Gordon86}
and in the non-linear/Gordon-Mollenauer phase noise\cite{Gordon90}(see for
example Ref.\cite{Mitra02}). The second feature worthy
of comment is the dependence of capacity upon length at large
distances. This is important, since it is for very long distance
communication that propagation non-linearities have the most
important effect. The Gaussian approximation to coherent
communication gives a lower bound on capacity that decays
exponentially with length at large distances. We do not expect
this exponential decay to correctly reproduce the dependence of
the system capacity, since it is in this limit that non-Gaussian
corrections to the input/output statistics are likely to be most
important. We will see below how taking account of these
non-Gaussian corrections is likely to change the dependence of
capacity at large distances.

Unfortunately, there exist no rigorous bounds on the capacity for
incoherent communication in which the phase of the electric field
is ignored. Nevertheless, expressionssimilar to 
Eq.(\ref{coherent_capacity}) can provide useful estimates of the incoherent
capacity. One such estimate is obtained by fitting an approximate
Gaussian distribution of intensity about its mean values (given by
Eq.(\ref{quadratic_stats})) to the quadratic statistics of
intensity given by Eq.(\ref{quartic_stats}). The capacity of the
channel with this Gaussian distribution is an approximation to the
capacity of the actual channel. It is given by
\begin{eqnarray}
\lefteqn{
c \approx
 \frac{1}{2} \int \frac{d \omega}{2 \pi \Delta} } & &
\nonumber\\
& &
\times
\log \left(
1
+ \frac{ P^2 2 \pi | {\cal G}^{II}_d(\omega) |^2/\Delta }{
P^2\left(1-2 \pi |  {\cal G}^{II}_d(\omega) |^2/\Delta \right)
+2PN+N^2 } \right). 
\label{C2}
\end{eqnarray}
The pre-factor of $1/2$ arises because the intensity is a real
field. In the limit of strong non-linearity and large signal to
noise ratio, Eq.(\ref{C2}) reduces to $C\approx  \sqrt{\pi/2}
\Delta_{eff} \hbox{Li}_{3/2} \left[P^2/(P+N)^2 \right]$. This has
an appealing interpretation. The effects of non-linearity are
contained entirely within a prefactor and provide an effective
bandwidth $\Delta_{eff}= 6\Delta \sqrt{L_D^2/ \eta L_{tot}^3}$.
Our analysis of jitter showed that XPM mainly introduces timing errors
and not amplitude error. The effective bandwidth enters because
signal pulses must be separated by more than the timing error in
order to be distinguishable. The information carried by each pulse
is determined by the additive noise power. Notice that in the
limit of strong non-linearity, the coherent capacity has an
exponential decay with length, whereas Eq.(\ref{C2}) has a
power-law decay. The actual coherent capacity is greater than the
incoherent capacity, since a greater number of degrees of freedom are
used in coherent communication. The behavior of
Eqs.~(\ref{coherent_capacity}) and (\ref{C2}) at large powers point to
limitations of the Gaussian approximation.
A comparison of the coherent and incoherent capacity is shown in
Fig.5. The incoherent capacity shows a peak at some optimal
power. This peak is at slightly higher power than that for the
coherent capacity, because phase/timing fluctuations are more
sensitive to XPM than amplitude fluctuations.

It seems plausible that it should be possible to express the 
capacity of a channel purely in terms of its multiple order 
Green's functions. Unfortunately, we do not have such an 
expression. Indeed, only for a channel that is entirely 
determined by its two-point (or single particle) Green's function 
is it possible to write down such an expression. In this case the 
channel is Gaussian and the result is Eq.(\ref{coherent_capacity}) 
as obtained by Shannon\cite{Shannon48}. Obtaining 
such expressions would be a significant contribution to the 
understanding of communication in non-linear media. The bound on 
coherent communication and the estimate of capacity for incoherent 
communication given above give an indication of the importance of 
including higher order Green's functions. Vertex corrections to 
higher order Green's functions determine the extent to which the 
channel has non-Gaussian statistics. Including these corrections 
can dramatically modify the functional dependence of ones 
estimates of system performance upon system parameters.

\begin{figure}
\centering
\epsfig{file=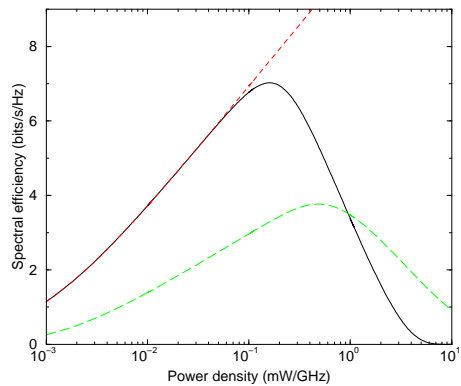,height=2in}
\caption{Spectral Efficiency {\it vs} Input Power: 
Plots of 
i. Eq.(\ref{Gaussian_Capacity})-solid black line;
ii.Shannon
formula, $c=\log [1+P/N]$ -short-dashed, red line;
iii.Eq.(29)-
long-dashed, green line.}
\end{figure}

\section{Conclusions}

In conclusion, we have considered the impediments to 
communication caused by non-linear interaction between the 
sub-bands of a WDM optical fiber system. These interactions 
between sub-bands lead to non-Gaussian input/output statistics. 
We have provided a simple characterization of these non-Gaussian 
statistics in terms of vertex corrections to higher order Green's 
functions of the channel. Propagation of light in the WDM system 
under the action of XPM is described by a Schr\"odinger equation 
with a spatially and temporally random potential. This potential 
encodes the effects of XPM. We have calculated the Green's 
function for propagation in this channel using Feynman 
path-integral and diagrammatic techniques. In the case where the 
signal correlations are approximated by delta-functions in time, 
these Green's functions may be obtained in closed form. The 
result shows a super-diffusive spreading of a signal pulse and of 
its arrival time. We have interpreted these results in terms of the
amplitude and timing jitter of received pulses; both of which are 
standard measures of optical system performance. We find that the 
received signal pulse shows a super-diffusive spreading of its 
distribution of arrival times; $\langle t^2 \rangle \propto L^3$. 
Finally, we have used the results to obtain lower bounds and 
estimates of the capacity of the WDM system. These are based upon 
the Shannon result for the capacity of a Gaussian channel. In the 
case of coherent communication, they lead to a strict lower bound 
upon the capacity. In the case of incoherent communication, where 
only the signal amplitude and not its phase is used for 
communication, we obtain an estimate of the system capacity. The 
different dependences of our estimates of capacity for coherent 
and incoherent communication at large distances, exponential and 
power-law, respectively, show the importance of accounting for 
the non-Gaussian nature of the channel.

\begin{appendix}
\section{Calculation of Amplitude Fluctuations}
The results contained in Eqs.(22) may be derived using rules for the
composition,
\begin{equation}
\int dt' {\cal G}(x_1,x';t_1,t'){\cal G}(x',x_2;t',t_2)
=
{\cal G}(x_1,x_2;t_1,t_2),
\label{A1}
\end{equation}
unitarity,
\begin{equation}
{\cal G}(x=0;t_1,t_2)
=
\delta (t_1-t_2)
\label{A2}
\end{equation}
and time-reversal,
\begin{equation}
{\cal G}^*(x_1,x_2;t_1,t_2)
=
{\cal G}(x_2,x_1;t_2,t_1)
\label{A3}
\end{equation}
of the Green's function for a particular realization of the potential,
$V(x,t)$. These properties of the Green's function may be confirmed by
considering the path integral formulation, Eq.(6). 

Using these relations, the first of Eqs.(22) may be deduced by the
following series of manipulations:
\begin{eqnarray*}
\langle
E_{out}(t)
\rangle
&=&
\langle
A\int dt' {\cal G}(t,t') +n(t)
\rangle
\\
&=&
A
\langle 
\int dt' {\cal G}(t,t')
\rangle
\\
&=&
A
\langle
{\cal G}(\omega=0)
\rangle
\end{eqnarray*}
This is nothing more than the average of Eq.(3). The second of
Eqs.(22) requires a few more manipulations;
\begin{eqnarray*}
\langle
|E_{out}(t)|^2
\rangle
&=&
A^2
\langle
\int dt_1 dt_2
{\cal G}(t,t_1)
{\cal G}^*(t,t_2)
\rangle
+
\langle |n(t)|^2 \rangle
\\
& &
+
A
\langle
n(t)
\int dt_1
{\cal G}^*(t,t_1)
\rangle
+
A
\langle
n(t)^*
\int dt_1
{\cal G}(t,t_1)
\rangle
\\
&=&
A^2
\int dt dt_1 dt_2
{\cal G}(t,t_1) {\cal G}^*(t,t_2)/\delta(0)
+N\delta (0)
\\
&=&
A^2
\int dt dt_1 dt_2
{\cal G}(0,0;t_2,t_1)/\delta(0)
+N\delta (0)
\\
&=&
A^2
\int dt dt_1 dt_2
\delta (t_1-t_2)/\delta (0)
+
N \delta(0)
\\
&=&
(A^2+N) \delta(0)
\end{eqnarray*}
We have used Eq.(\ref{A3}) in moving from the first to second line,
followed
by Eqs.(\ref{A1}) and (\ref{A2})  moving between lines two and three
and three and four, respectively. We have also used the fact that the
averages over the noise and over the potential are temporally invariant.
Eq.(22c) may be deduced as follows:
\begin{widetext}
\begin{eqnarray*}
\langle
\delta |E_{out}(t)|^2
\delta |E_{out}(t)|^2
\rangle
&=&
\langle
|E_{out}(t)|^2
|E_{out}(t)|^2
\rangle
-
\langle
|E_{out}(t)|^2
\rangle^2
\\
&=&
A^4\int dt_1 dt_2 dt_3 dt_4
\langle
{\cal G}^*(L;t,t_1)
{\cal G}^*(L;t,t_2)
{\cal G}(L;t,t_3)
{\cal G}(L;t,t_4)
\rangle
\\
& &
+
4N
A^2
\int dt_1 dt_2
\langle
{\cal G}*(L;t,t_1)
{\cal G}(L;t,t_2)
\rangle
+
\left[2N^2 -(A^2+N)^2 \right]\delta(0)^2
\\
&\approx&
A^4\int dt_1 dt_2 dt_3 dt_4
\langle
{\cal G}^*(L;t,t_1)
{\cal G}(L;t,t_3)
\rangle
\langle
{\cal G}^*(L;t,t_2)
{\cal G}(L;t,t_4)
\rangle
+
\left[N(2A^2+N)-A^4\right] \delta(0)^2
\\
&\approx&
A^4
\left[\int dt dt_1 dt_2 
\langle
{\cal G}^*(L;t,t_1)
{\cal G}(L;t,t_3)
\rangle/\delta(0)
\right]^2
+
\left[N(2A^2+N)-A^4\right] \delta(0)^2
\\
&\approx&
N(2A^2+N) \delta(0)^2
\end{eqnarray*}
\end{widetext}
The third line is simply a rearrangement of the various terms arising
after substitution of Eq.(3). A complete calculation of the first term
in this expression requires evaluation of the 4-particle Green's
function. We do not carry out this calculation here. Instead we make
the approximation discussed in the text and ignore vertex
corrections. This approximation is embodied in the step between the
third and fourth lines. In moving to the fifth line, we have used the
fact that the average over the noise and random potential leads to
temporally invariant expressions. We may then
use Eqs.(\ref{A1},\ref{A2}) and (\ref{A3}) in the same way as before
in order to derive our final result.
\end{appendix}


\end{document}